\documentclass[%
 aip,
 jap,%
 amsmath,amssymb,
 reprint,%
]{revtex4-1}

\usepackage{graphicx}
\usepackage{dcolumn}
\usepackage{bm}

\begin{document}
\title{Surface potential at a ferroelectric grain due to asymmetric
screening of depolarization fields}
\author{Yuri A. Genenko} \email{genenko@mm.tu-darmstadt.de} %
\author{Ofer Hirsch}
\altaffiliation{New address: Swiss Federal Institute of Technology, Z\"urich, Switzerland}
\affiliation{Technische Universit\"at Darmstadt, %
Darmstadt, Germany}%
\author{Paul Erhart}
\affiliation{Chalmers University of Technology, Gothenburg, Sweden}

\date{\today}

\begin{abstract}
Nonlinear screening of electric depolarization fields, generated by a stripe domain structure in a ferroelectric grain
of a polycrystalline material, is studied within a semiconductor model of ferroelectrics. It is shown that the maximum
strength of local depolarization fields is rather determined by the electronic band gap than by the spontaneous polarization
magnitude. Furthermore, field screening due to electronic band bending and due to presence of intrinsic defects leads 
to asymmetric space charge regions near the grain boundary, which produce an effective dipole layer at the surface of 
the grain. This results in the formation of a potential difference between the grain surface and its interior of the 
order of $1 \rm\: V$, which can be of either sign depending on defect transition levels and concentrations. Exemplary 
acceptor doping of $\rm BaTiO_3$ is shown to allow tuning of the said surface potential in the region between $0.1$ 
and $1.3 \rm\: V$. 
\end{abstract}

\maketitle

\section{\label{sec:intro}Introduction}

Potential barriers at internal interfaces of polycrystalline materials have a great impact on their physical
properties, particularly, on dielectric properties and nonlinear ionic and electronic conductivity 
\cite{Johnson1999APL,Molak2008,Froemling2012,Andrejs2013}. A physical reason for the formation of the barriers 
is often redistribution of charged defects at grain boundaries. In case of conducting oxides it is the segregation 
of oxygen vacancies, the most mobile charge defects, that form - together with immobile background ions - space 
charge regions resulting in the electrostatic potential barriers \cite{Nyman2012APL,Lee2012afm}. 

In the special case of ferroelectric ceramics potential barriers may result from the spontaneous polarization and 
consequent internal depolarization fields which do not vanish entirely in a disordered medium. Local depolarization
fields have a strong impact on formation of polarization structures in ferroelectrics
\cite{Kittel1946,LandauElectrodynamicsContinuum,Mitsui1953}. They can also trigger charge defect migration 
which is considered as a possible factor of aging and fatigue of ferroelectrics 
\cite{Mitsui1953,takahashi70space,thomann72stabilization,lupascu06aging,Genenko-agingPRB2007,GenenkoPRB2008,BalkeJAP2009}
affecting performance of these materials used in sensors, actuators and non-volatile random-access memory devices. The
magnitude of these electric fields produced by bound charges due to spontaneous polarization may be remarkable in 
comparison with coercive fields (1-10 $\rm\: kV/mm$), however, observation of these fields is difficult since they reveal
themselves only at the micro- to mesocale. Nevertheless, recent measurements of the electric potential on the surface of
barium titanate single crystals by using ultrahigh-vacuum atomic force microscopy have shown periodic step-like potential
structures typical of upward and downward $180^{\circ}$ domains in this material \cite{WatanabeFerro2008,Kaku2009}. On the
other hand, the amplitude of the potential variation appeared to be two orders of the magnitude smaller than that predicted
by the classical theory of a stripe domain structure \cite{LandauElectrodynamicsContinuum,Fedosov1976}.

Drastic differences between experiment and this simple model of a ferroelectric were supposed to result from the 
distortion of the electronic band structure by the electric field \cite{Kaku2009}. Indeed, variations of the electrostatic
potential at the scale of a typical domain width in barium titanate may amount to several volts, while the band gap in this
compound is about $3.4\,\rm\: eV$. Therefore the material has to be considered as a wide-gap semiconductor 
\cite{Fridkin1980,WatanabePRB1998}. Band bending near the positively charged domain boundaries leads thus to 
formation of space 
charge regions with an excessive electron concentration, while band bending near the negatively charged domain boundaries
creates space charge regions with an excessive hole concentration. In both cases this results in the depression of the
electric field which causes the band bending. Hence, the distribution of charges and fields is governed by the
self-consistent nonlinear Poisson equation accounting for the electronic band structure of the material.

Beside electronic carriers a significant contribution to the field screening can be made by various charged
defects in ferroelectric perovskites which are typically vacancies and - intentional or
unintentional - impurities. Their contribution to charge balance and the formation of space charge regions depends on the
position of the defect energy levels with respect to the band edges as well as their concentration. This allows in 
principle a fine control of the screening of the depolarization field and related physical properties by doping
ferroelectrics with certain donor or acceptor impurities or their combinations. This understanding was confirmed by recent
investigations of the photochemical reactions with a variety of metal salts on a surface of the lead zirconate titanate 
where the variation of the conduction band edge of about $\pm 0.5\,\rm\: V$ depending on the local polarization state was
established \cite{Jones2008}.

So far, a thorough quantitative analysis of the nonlinear electric field screening was performed only in the 
one-dimensional case \cite{WatanabePRB1998,Gureev2011} or for a single domain wall in the film geometry 
\cite{Xiao2005} which misses some important features and consequences of the screening in the case of domain arrays in 
the bulk material. In this work the depolarization field problem is treated in a two-dimensional model of a ferroelectric 
grain \cite{Genenko-agingPRB2007} extended by the above-mentioned nonlinear Poisson equation. The model furthermore involves 
the evaluation of intrinsic defect concentrations from thermodynamic balance equations \cite{ErhartJAP2008} using defect 
transition levels calculated from density functional theory (DFT) \cite{ErhartJAP2007}. The paper is organized as follows. 
In Section \ref{sec:nonlinearlmodel} 
a nonlinear semiconductor model of a ferroelectric grain is formulated including the nonlinear Poisson equation and 
evaluation of the charge defect densities. Numerical solution of the semiconductor model by means of the finite-element (FE) 
method is delineated in Section \ref{sec:intrinsic} for the case of only intrinsic defects present. Effect of extrinsic 
doping on charge and potential distributions is studied in Section \ref{sec:doping}. Physical results of the nonlinear 
field screening in differently doped ferroelectrics are finally concluded in Section \ref{sec:conclusions}. 
In Appendices, Green's function of a linear anisotropic problem is derived which is used for verification of the nonlinear
numerical calculations in Section \ref{sec:intrinsic}.

\section{\label{sec:nonlinearlmodel} Semiconductor model of a ferroelectric grain}

\begin{figure}[t]
\begin{center}
    \includegraphics[width=7cm]{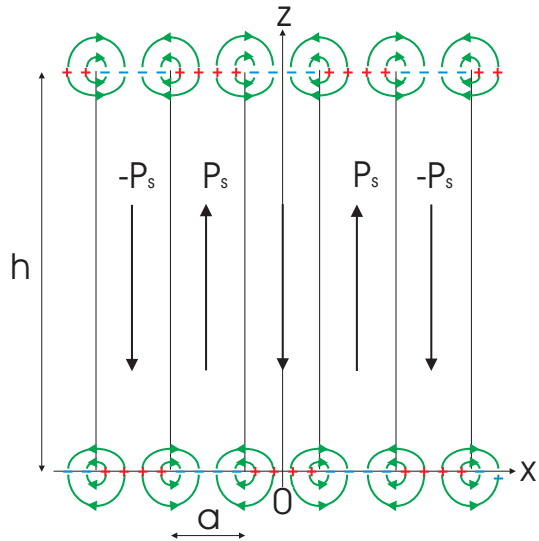}
    \caption{(Color online) Layout of a 2D-array of $180^{\circ}$-domain walls crossing the grain boundaries
     at a right angle. Straight arrows show the direction of the polarization and
     curved arrows the schematic pattern of the local electric fields.}
    \label{domarray}
\end{center}
\end{figure}

In this section the main components of the nonlinear electrostatic model are presented: the model geometry, governing
equations and boundary conditions. Our consideration is based on the two-dimensional model of an isolated ferroelectric 
grain inside an unpoled polycrystalline ferroelectric suggested in \cite{Genenko-agingPRB2007,GenenkoFerro2008}
which applies, in fact, to any poly-domain single crystalline sample electrically decoupled from surrounding. The
quadratic grain of size $h$ is filled with an array of stripe domains of width $a\ll h$ as is schematically shown in 
Fig.~\ref{domarray}. The full polarization of the grain equals zero. A hard domain structure is assumed, i.e. the spatial
variation of the polarization within the domains is neglected as is appropriate by temperatures well below the
ferroelectric phase transition temperature. Since depolarization fields created by bound charges at the grain boundary
exponentially decay on the typical distance of $a$ \cite{Genenko-agingPRB2007} the grain separated from the other grains 
by a dielectric layer of comparable thickness may be considered as electrically decoupled from the surrounding. For the same
reason, by evaluation of the electric field it suffices to consider just one side of the quadratic frame. Furthermore, 
FE computations of the field in such a frame show that the field pattern is virtually periodic with the exception of the 
very edges of the domain array as soon as $a\ll h$ \cite{GenenkoFerro2008}. 
That is why in the following numerical treatment we will study just one repetitive element of a two-dimensional periodic
array of domains infinite in the $x-$direction and cut by the surfaces $z=0$ and $z=h$ perpendicular to the direction of 
spontaneous polarization in domains.

The ferroelectric medium occupies the region $0<z<h$ and is characterized by the tensor of dielectric permittivity
$\hat\varepsilon=\varepsilon_0 \hat\varepsilon_f$ with $\varepsilon_0$ the permittivity of vacuum, which is assumed to 
be diagonal in the chosen Cartesian frame:
\begin{equation}
\hat\varepsilon_f= 
\left(
\begin{array}{lcr}
\varepsilon_a & 0 & 0 \\
0 & \varepsilon_b & 0 \\
  0 & 0 & \varepsilon_c
\end{array}
\right)
\label{etensor}
\end{equation}
The semi-spaces $z<0$ and $z>h$ are occupied by an isotropic dielectric medium characterized by the relative 
dielectric constant
$\varepsilon_d$. The system is supposed to be uniform in the $y-$direction so that no quantities involved are $y$ dependent.
This model configuration is well-known in the physics of polarized media and was used for the study of equilibrium and
dynamic properties of ferromagnetic \cite{Kittel1946,LandauElectrodynamicsContinuum} and ferroelectric 
\cite{Mitsui1953,Fedosov1976} materials.

Due to the spontaneous polarization $\mathbf{P}_s$, the domain faces at $z=0$ and $z=h$ are alternatively charged 
with the bound surface charge density $\sigma = |\mathbf{P}_s|$. The electric field ${\bf E}(x,z)$ is determined by 
the bound surface charge and the total space charge $\rho(x,z)$ of free carriers and charged defects 
through Gauss' law
\begin{equation}
\label{Gauss}%
\nabla(\hat\varepsilon_f{\bf E})=\rho(x,z).
\end{equation}
Assuming the total electroneutrality of the system and the same periodicity of $\rho(x,z)$ along the $x$ axis 
as that of the domain array the electric field has to vanish far away from the grain boundaries $z=0$ and $z=h$ that serves 
as the asymptotic boundary condition for the electric field. Natural boundary conditions at the grain boundaries
 are given by continuity of the electrostatic potential $\varphi$ and of the 
normal electric displacement component at the boundaries $z=0$ and $z=h$ \cite{LandauElectrodynamicsContinuum}.

\subsection{\label{subsec:nolineq} Constitutive equations }

Distributions of the electrostatic potential $\varphi (x,z)$ in the ferroelectric and the dielectric regions obey the
Poisson equation (\ref{Gauss}) where the charge density $\rho$ on the right-hand side includes all the charged 
species relevant for undoped $\rm BaTiO_3$ synthesized under Ba-rich conditions\cite{ErhartJAP2007}:
\begin{equation}
\label{rho}
\rho=q\left( p - n + 2N_{V_{O}^{2+}} - 4N_{V_{Ti}^{4-}} - 2N_{\left[V_{Ti}-V_O\right]^{2-}}                             
\right).
\end{equation}
\noindent
Here $q$ denotes the elementary charge, $p$ and $n$ the densities of holes and electrons, respectively, and 
$N_{V_{O}^{2+}},\,N_{V_{Ti}^{4-}}$ and $N_{\left[V_{Ti}-V_O\right]^{2-}}$ 
the densities of the respective ionized defects in the indicated charged states. Note that we assume the defect
concentrations $N_{V_{O}}, N_{V_{Ti}}, N_{\left[V_{Ti}-V_O\right]}$  to be homogeneous over the entire sample and 
thus neglect possible segregation effects that have been shown to exist e.g., in $\rm BaZrO_3$ \cite{Nyman2012APL}.

All the particular charge densities are dependent on the local value of the electrostatic potential as follows
\cite{Sze}
\begin{align}
\label{n-density}
n &=  N_C \frac{2}{\sqrt{\pi}} F_{1/2} \left( \frac{ E_F-E_{CB}+q\varphi }{ k_B T} \right),\\
\label{p-density}
p &=  N_V \frac{2}{\sqrt{\pi}} F_{1/2} \left( \frac{E_{VB}-E_F-q\varphi }{k_B T} \right),\\
\label{VO-density} 
N_{V_{O}^{2+}} &=  \frac{ N_{V_{O}} } {1 + g_D\exp{\left( \frac{\displaystyle E_F-E_{V_{O}^{2+}}+q\varphi }
{\displaystyle k_B T} \right)}},\\
\label{Ti-density} 
N_{V_{Ti}^{4-}} &=  \frac{N_{V_{Ti}} } {1 + g_A\exp{\left( \frac{\displaystyle E_{V_{Ti}^{4-}}-E_F-q\varphi }
{\displaystyle k_B T} \right)}},\\
\label{TiO-density} 
N_{\left[V_{Ti}-V_O\right]^{2-}} &=  \frac{N_{\left[V_{Ti}-V_O\right]}} 
{1 + g_A\exp{\left( \frac{\displaystyle E_{\left[ V_{Ti}-V_O \right]^{2-}}-E_F-q\varphi }{\displaystyle k_B T} \right)} }.
\end{align}  
\noindent
where $k_B$ is the Boltzmann constant, $ T=300 \rm\: K$ absolute temperature, $N_C$ and $N_V$ the effective densities
of states in the conduction band and  in the valence band, respectively \cite{WechslerJOSA1988}, $F_{1/2}(x)$ the 
complete Fermi-Dirac integral \cite{Abramowitz}. The degeneracy of the defect level is set to two in the donor case 
($g_D=2$) and to four in the acceptor case ($g_A=4$) to account for the spin polarization of electrons and holes 
\cite{Sze}. The Fermi energy $E_F$ is defined far away from the charged interfaces at $z=h/2 >> a$ by setting the
electrostatic potential $\varphi $ and the right-hand side of Eq.~(\ref{rho}) to zero.

Depending on the defect energies and concentrations the densities of electrons and holes may be in certain 
circumstances rather large. Then the question arises whether redistribution of these mobile charge carriers can 
compensate the depolarization field completely. Conditions of equilibrium with regard to the drift and diffusion 
of electrons and holes can be formulated as vanishing currents of both species:    
\begin{align}
\label{currents=0}
j_{n} &= -q\mu_{n}n\nabla \varphi +qD_n\nabla n=0,\nonumber\\
j_{p} &= -q\mu_{p}p\nabla \varphi -qD_p\nabla p=0 
\end{align}
\noindent
where $\mu_{n}$ ($\mu_{p}$) and $D_n$ ($D_p$) are the mobility and diffusivity of electrons (holes), respectively. 
Since in our problem the Fermi energy may cross the valence and the conduction band edges the Fermi statistics should be 
used which makes the classical Einstein relation between diffusivity and mobility, $ \mu= qD/k_B T$, invalid. In this case, 
the generalized Einstein relations \cite{Ashkroft,GenenkoJAP2006} should be applied which read 
\begin{equation}
\label{Einstein}
\mu_{n}=qD_n \frac{1}{n} \frac{\partial n}{\partial E_F},\,\,\,
\mu_{p}=-qD_p \frac{1}{p} \frac{\partial p}{\partial E_F}.
\end{equation}
\noindent
With these relations implemented, Eqs.~(\ref{currents=0}) become compatible with equilibrium expressions for the 
charge carrier densities (\ref{n-density}) and (\ref{p-density}). This means, particularly, that the depolarization 
field can coexist with nonuniform charge carrier distributions at mesoscopic scale in equilibrium.

The system of equations 
(\ref{Gauss},\ref{rho},\ref{n-density},\ref{p-density},\ref{VO-density},\ref{Ti-density},\ref{TiO-density}) 
can be numerically solved as soon as the material parameters 
and concentration of defects are specified. The choice of the latter is detailed in the next section.

\subsection{\label{subsec:defects} Evaluation of the intrinsic defect concentrations }

Even in the nominally undoped $\rm BaTiO_3$ ceramics a number of defects appear during the sintering process at
high temperatures making the material intrinsically doped. 
\begin{table}[t]
\caption{\label{Materialparameters1} Material and model parameters}
\begin{center}
\begin{tabular}{p{2.4in}|r|}                                                       \hline
Band gap, $E_G$   & $3.4 \rm \: eV$                                       \\  
Transition level of oxygen vacancy, $E_{V_{O}^{2+}}$ & $3.35 \rm \: eV$  \\  
Transition level of titanium vacancy, $E_{V_{Ti}^{4-}}$ & $0.4 \rm \: eV$ \\  
Transition level of titanium-oxygen di-vacancy, $E_{\left[ V_{Ti}-V_O \right]^{2-}}$  & $0.21 \rm \: eV$  \\  
Oxygen vacancy density, $N_{V_{O}}$ & $1.214\times 10^{20} \rm \: m^{-3}$ \\  
Titanium vacancy density, $N_{V_{Ti}}$ & $8.494\times 10^{21} \rm \: m^{-3}$ \\  
Titanium-oxygen di-vacancy density, $N_{\left[V_{Ti}-V_O\right]}$ & $1.370\times 10^{22} \rm \: m^{-3}$ \\  
Density of states of the valence band, $N_{V}$ & $1.5\times 10^{28} \rm \: m^{-3}$ \\  
Density of states of the conduction band, $N_{C}$ & $1.6\times 10^{28} \rm \: m^{-3}$ \\  
Relative permittivity in crystallographic $\,$ direction $a$, $\varepsilon_a$ & $2180$ \\  
Relative permittivity in crystallographic $\,$ direction $c$, $\varepsilon_c$ & $56$ \\  
Relative permittivity of the dielectric, $\varepsilon_d$  & $1$  \\  
Spontaneous polarization in direction $c$, $P_s$  & $0.25 \rm \: C m^{-2}$  \\  
Domain width, $a$ & $100 \rm\: nm$ \\  
Domain length, $h$ & $40\, a$ \\  \hline 
\end{tabular}
\end{center}
\end{table}
The type and concentrations of defects strongly depend on conditions of the material synthesis resulting 
in a certain position within the stability diagram of the compound \cite{ErhartJAP2008,ErhartJAP2007}. 
Typical natural acceptor and donor defects, 
which form in $\rm BaTiO_3$ during the production procedure under Ba-rich conditions, are exemplarily considered here 
with respect to their role in field screening at grain boundaries. According to DFT calculations the 
most favorable defects are then doubly ionized oxygen vacancies, $V_{O}^{2+}$, which act as donors, as well as titanium 
vacancies $V_{Ti}^{4-}$ and di-vacancies $\left[V_{Ti}-V_O\right]^{2-}$, which both act as acceptors \cite{ErhartJAP2007}.
Their transition energy levels with respect to the top of the valence band are presented in Table~\ref{Materialparameters1}
together with other material and model parameters taken from Refs. \cite{WechslerJOSA1988} and \cite{Zgonik1994}.
Defect concentrations were calculated according to the procedure 
described in \cite{ErhartJAP2008} using defect formation energies from \cite{ErhartJAP2007}. It has been shown that this
approach yields defect concentrations and electrical conductivities in excellent agreement with experimental 
high-temperature data over a wide range of oxygen partial pressures \cite{ErhartJAP2008}. For our calculations the sample 
was assumed to be fully equilibrated at $T=1000 \rm \: K$ 
, at an atmospheric oxygen partial 
pressure of $0.21\times 10^5\rm \: Pa$, followed by rapid quenching to $300 \rm \: K$. The
concentrations of barium vacancies and barium-oxygen di-vacancies as well as defects $V_{O}$, $V_{Ti}$ and $\left[V_{Ti}-V_O\right]$ in other ionization states than those shown in Table~\ref{Materialparameters1} 
(for example, single- and double- ionized Ti vacancies) are orders of the magnitude smaller than 
$N_{V_{O}^{2+}}$, $N_{V_{Ti}^{4-}}$ and $N_{\left[V_{Ti}-V_O\right]^{2-}}$, and thus have been neglected.

\subsection{\label{subsec:semi-geometry} Framework and boundary conditions for numerical calculations}

\begin{figure}[t]
\begin{center}
    \includegraphics[width=6.0cm]{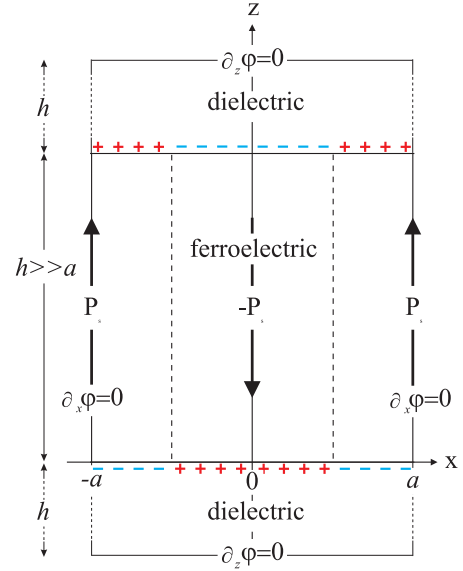}
    \caption{(Color online) The layout of the computational framework with boundary conditions indicated. }
    \label{comp-area}
\end{center}
\end{figure}

Due to periodicity, it is sufficient to consider just one repetitive element of the 
structure shown in Fig.~\ref{domarray}. For the numerical treatment, we choose the area $-a < x <a,\, -h < z < 2h$
with $h=40\,a$ including two domain walls at the positions $x=\pm a/2$ which separate adjacent positively and negatively 
charged domain boundaries, as shown in Fig.~\ref{comp-area}. The ferroelectric material occupies the area $0< z< h$ 
while the external regions, $-h < z < 0$ and $h < z < 2h$, are occupied by the dielectric. The boundaries of each domain 
at $z=0$ and $z=h$ are charged with the surface charge density $\pm \sigma$ as is shown in Fig.~\ref{comp-area}. Thus, 
in the middle of the frame, $\vert x\vert < a/2$, polarization is negative while in the outer regions, 
$a/2 < \vert x\vert < a$, polarization is positive.

The following requirements are used as the boundary conditions: (a) the electric field vanishes far away from the charged
domain boundaries; for the chosen computational framework this means $\partial_z\varphi = 0$ at $z = 2h$ and at $z = -h$; 
(b) since the periodic domain structure is bilaterally symmetrical with respect to the centers of both positive and 
negative domains, the transverse field component vanishes at the side boundaries of the computational framework,
$\partial_x\varphi = 0$ at $x =\pm a$; (c) for the charged boundaries at $z = 0$ and $z = h$ the natural boundary 
conditions apply which follow from Gauss' law \cite{jackson75classical}, 
\begin{align}
\label{BCframetop}
\varepsilon_d \partial_z \varphi (x,h+0) - \varepsilon_c \partial_z \varphi (x,h-0) 
& =  -\sigma_p(x,h),\\
\label{BCframebot}
\varepsilon_c \partial_z \varphi (x,+0) - \varepsilon_d \partial_z \varphi (x,-0) 
& =  -\sigma_p(x,0) 
\end{align}  
where the local values of the surface charge densities at the ferroelectric boundaries, $\sigma_p(x,z)$,
adopt constant values $\pm\sigma $ as indicated in Fig.~\ref{comp-area}. Note that, in principle, the surface
charges may be included either in the right-hand side of Eq.~(\ref{Gauss}) as $\delta$-functions, or in the 
boundary conditions. For implementation of the FE calculations the second approach is adequate using the boundary 
conditions (\ref{BCframetop},\ref{BCframebot}).

\section{\label{sec:intrinsic} Charge and potential distributions in the intrinsically doped $\rm BaTiO_3$}

\subsection{\label{sec:FEresults} FE evaluation of charge and potential profiles}

\begin{figure}[t]
\begin{center}
    \includegraphics[width=8.0cm]{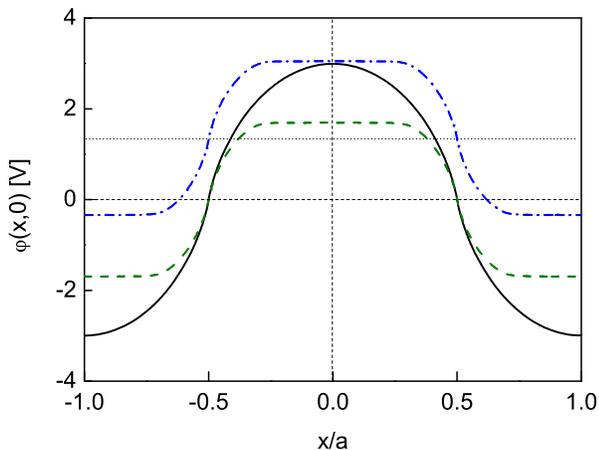}
    \caption{(Color online) Electrostatic potential profile in $x$ direction along the ferroelectric/dielectric 
             interface at $z = 0$. Solid and short-dashed lines show the numerical and the analytical calculation 
             in absence of free charges, respectively (the curves cannot be disnguished). The dashed line presents 
             the potential with account of the electronic charges $p$ and $n$ only while the dash-dotted line 
             accounts for both electronic and defect charges, the thin dotted horizontal line indicating its mean 
             value $\varphi_s$.}
    \label{phix-profile}
\end{center}
\end{figure}

\begin{figure}[t]
\includegraphics[width=8.0cm]{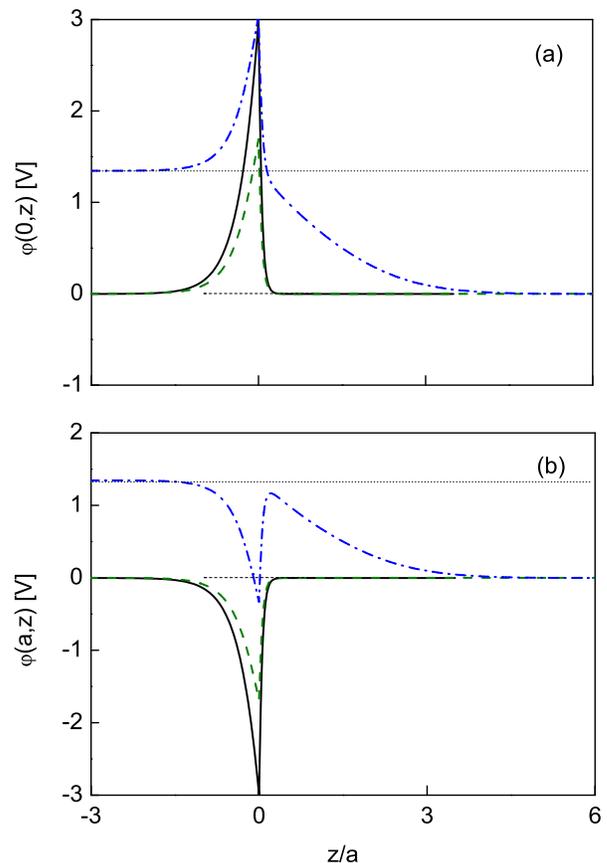}
\caption{\label{phiz-profile} (Color online)
Electrostatic potential profiles in $z$ direction along the domain symmetry axes at $x = 0$ (a) and $x = a$ (b).
Solid lines show the potential distributions in absence of free charges, the dashed line in the presence of the 
electronic charges only and the dash-dotted line with account of both the electronic and the defect charges. 
The thin dotted horizontal lines indicate the mean value $\varphi_s$ at the surface.}
\end{figure}

The system of equations 
(\ref{Gauss},\ref{rho},\ref{n-density},\ref{p-density},\ref{VO-density},\ref{Ti-density},\ref{TiO-density}) 
with the input parameters from the Table~\ref{Materialparameters1} has been solved using the FE software FlexPDE 
on the two-dimensional frame of Fig.~\ref{comp-area}. Results are presented exemplarily in Figs.~\ref{phix-profile} 
and ~\ref{phiz-profile} to illustrate the main features of the potential profile. 
To validate the numerical treatment 
the electrostatic potential at the charged interface $z=0$ was first calculated in absence of electronic and defect 
charges (solid line in Fig.~\ref{phix-profile}) and compared with the respective analytic result (short-dashed 
line in Fig.~\ref{phix-profile}) given by the formula
\begin{equation}
\label{Phi_x-anal}
\varphi_b(x,0)=\int_{z}^{\infty} dz\, E_z^0 (x,z)
\end{equation}
\noindent where the field is defined by Eq.~(\ref{E0z+}) of Appendix B.
These two lines coincide perfectly and present periodic alternating variation of the potential with a maximum about 
$3 \rm\: V$ in the middle of the positively charged domain boundary ($x=0$) and a minimum of the same magnitude but 
negative sign in the middle of the negatively charged domain boundaries ($x=\pm a)$. 

The dashed line represents the solution in the presence of electronic charge carriers only, i.e. in the limit 
that all defect densities $N_{V_{O}}, N_{V_{Ti}}, N_{\left[V_{Ti}-V_O\right]}$ in Eq.~(\ref{rho}) are set to zero. 
This was done to illustrate the pure effect of the electronic band bending alone. As well as the solid line this 
solution exhibits 
symmetry with respect to positively and negatively charged domain faces but with the magnitude of the alternating 
potential reduced to about $1.7 \rm\: V$. The plus and minus potential amplitudes remain symmetric because of the 
virtually equal parameters of the conduction ($N_C$) and the valence ($N_V$) bands \cite{WechslerJOSA1988} (see 
Table \ref{Materialparameters1}). Thus, account of the electronic band structure limits the maximum variation of 
the electrostatic potential to the band gap magnitude of $3.4 \rm\: V$. Stronger variations of the potential and, 
respectively, stronger electric fields are compensated by accumulation of the electronic carriers at the charged 
domain boundaries irrespectively of the magnitude of the spontaneous polarization $P_s$.

Finally the dash-dotted line represents the solution when both electronic carriers and charged defects are included.
In this case the symmetry between the positively and negatively charged domain boundaries is distinctly broken so that 
a mean value of the potential $\varphi_s = 1.34 \rm\: V$ prevails at the interface. The potential distribution looks
symmetrically alternating around $\varphi_s$ with an amplitude of $1.7 \rm\: V$.

For better understanding of the nature of the potential shift $\varphi_s$ the potential profiles along the symmetry 
axes of the positive and the negative polarization domains are plotted in Fig.~\ref{phiz-profile}. 
In the absence of both charge carriers and defects, potential peaks are due to positive and negative surface bound 
charges only (solid lines). When band bending is taken into account the potential peaks are reduced by approximately 
one half due to space charges of electrons and holes (dashed lines). Finally, in the presence of both free carriers and 
charged defects (dash-dotted lines), the asymptotic potential values to the left and to the right of the interface 
become different revealing a potential step along the $z$ direction. The average value of the potential at the 
interface with respect to the interior of the grain, becomes positive and equals $1.34 \rm\: V$ in accordance with 
Fig.~\ref{phix-profile}.

\begin{figure}[t]
\includegraphics[width=8.0cm]{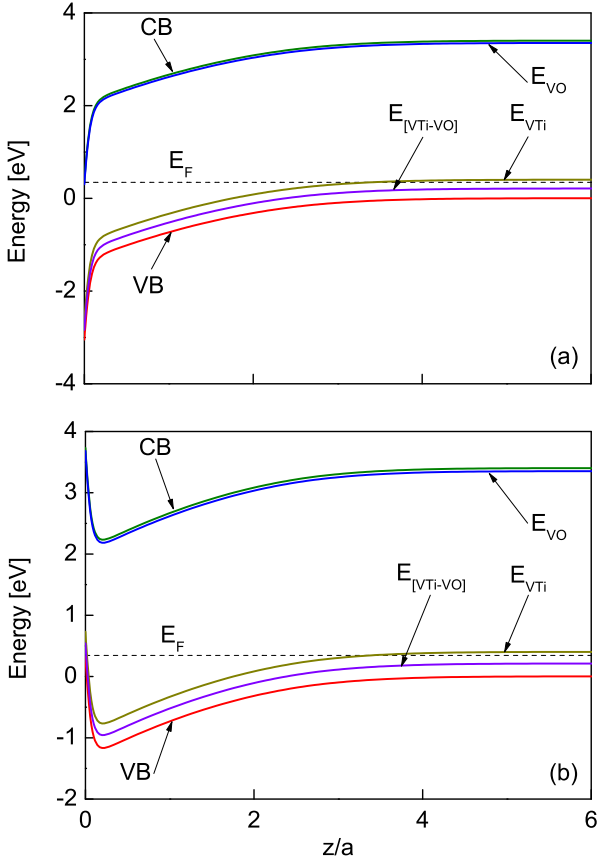}
\caption{\label{bands}(Color online)
Spatial variation of band edges and defect energy levels along the symmetry axis $x = 0$ of the negative
polarization domain (a) and along the symmetry axis $x = a$ of the positive polarization domain (b). 
$E_F = 0.344 \rm\: eV$}
\end{figure}
The spatial variations of conduction and valence band edges as well as defect transition levels are shown in 
Fig.~\ref{bands}. Similar to Fig.~\ref{phiz-profile} it discloses two characteristic length scales of the potential 
variation. The first one is intrinsic to the stripe domain structure and is about 
$a_s=a\sqrt{\varepsilon_c/\varepsilon_a} < a$. The second one, which arises only in the presence of the charged defects, 
is one order of magnitude larger and amounts to a few $a$. Though equations 
(\ref{Gauss},\ref{rho},\ref{n-density},\ref{p-density},\ref{VO-density},\ref{Ti-density},\ref{TiO-density}) 
are nonlinear the potential profiles in Fig.~\ref{phiz-profile} can be roughly interpreted as a superposition 
of the (screened) short-range potential due to the charged domain boundaries and the long-range potential step
across the ferroelectric/dielectric interface.

\begin{figure}[t]
\begin{center}
    \includegraphics[width=8.0cm]{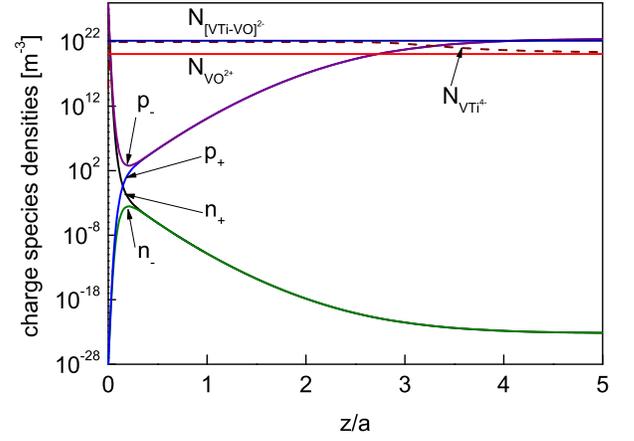}
    \caption{(Color online) Spatial distributions of charged species along the symmetry axes of domains in direction 
    $z$. Densities of electrons and holes are indicated as $n_{-}$ and $p_{-}$, respectively, in front of the 
    negatively charged boundary at $x=a$ and as $n_{+}$ and $p_{+}$ in front of the positively charged boundary at $x=0$.
    Concentrations of charged defects $N_{V_{O}^{2+}}$, $N_{\left[V_{Ti}-V_O\right]^{2-}}$ (solid lines) and 
    $N_{V_{Ti}^{4-}}$ (dashed line) are virtually independent on the charge at the boundaries except for the very narrow
    space region which is not visible in the picture.}
    \label{charge-profiles}
\end{center}
\end{figure}

Spatial distributions of charge carriers and charge defects corresponding to the potential distribution are
presented in Fig.~\ref{charge-profiles}. The densities of the charged defects $N_{V_{O}^{2+}}$ and
$N_{\left[V_{Ti}-V_O\right]^{2-}}$ remain virtually constant all over the system except for the regions of a few $\rm nm$
near the charged boundaries not seen in the figure. The density of the charged defects $N_{V_{Ti}^{4-}}$ in contrast
undergoes spatial variation at the same scale of about $5a$ as the charge carrier densities. The density of electrons 
thereby remains very small everywhere but the narrow region of about $0.2a$ in front of the positive boundary. The density 
of holes is in contrast high, particularly far away from the boundaries, to outweigh the high density of the negatively
charged defects. Note that the densities of all involved charged species resulting from continuous Eqs.~(\ref{n-density}) 
and (\ref{p-density})and displayed for completeness in the whole calculation domain 
in Fig.~\ref{charge-profiles} are unphysically small from the atomistic point of view in some regions. This concerns
particularly the densities of electrons $n_{-}$ (in the whole domain), $n_{+}$ (in the whole domain but the close 
vicinity of the positive boundary) and of holes $p_{+}$ (in the close vicinity of the positive boundary). This means 
that these species can be simply neglected in respective areas.   

\subsection{\label{sec:surf-dip} Appearance of a surface dipole layer}

The long-range contribution to the potential may appear if a surface dipole layer is present at the interface 
as it is the case in a deliberately doped p-n junction between two semiconductors \cite{Sze}. Then the mean value
$p_s$ of the surface dipole density, $p_z(x)$, can be easily related to the mean surface value of the
potential, $\varphi_s$. Indeed, 
\begin{equation}
\begin{split}
\label{p_s-phi_s}
p_s &= \frac{1}{2a}\int_{-a}^{a}dx\,p_z(x)=\frac{1}{2a}\int_{-a}^{a}dx\,\int_0^{\infty}dz\,z\,\rho(x,z)\\
&=-\frac{\varepsilon_0\varepsilon_c}{2a}\int_{-a}^{a}dx\,\varphi_(x,0)=-\varepsilon_0\varepsilon_c \varphi_s
\end{split}
\end{equation}
\noindent 
where Eq.~(\ref{Gauss}) and the corresponding boundary conditions from Section \ref{subsec:semi-geometry} 
were utilized.

\begin{figure}[t]
\begin{center}
    \includegraphics[width=6.7cm]{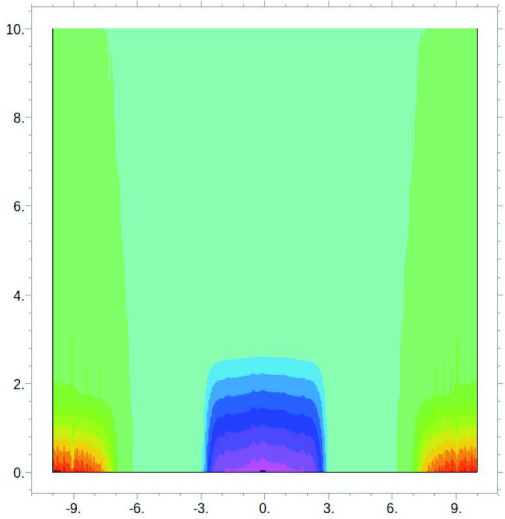}
    \caption{(Color online) FE calculation of the space charge distribution in front of the charged interface 
             $z =0$. The space charge density varies between the negative maximum of $-1.6\times 10^9 \rm\: C/m^{-3}$
             (dark) and the positive maximum of $1.8\times 10^9 \rm\: C/m^{-3}$ (bright). The vertical length 
             scale is in units of $10^{-10} \rm\: m$, the horizontal one is in units of $10^{-8} \rm\: m$.}
    \label{Flex}
\end{center}
\end{figure}

What can be a reason for the formation of the effective dipole density at the ferroelectric/dielectric interface? 
To comprehend this phenomenon the charge distribution obtained by FE calculations is displayed in Fig.~\ref{Flex}. 
Space charge regions of different extensions are clearly seen in front of the positively and negatively charged 
parts of the interface. They result in unbalanced contributions to the dipole density and consequently 
to the nonzero mean potential at the interface. 
\begin{figure}[t]
\begin{center}
    \includegraphics[width=7.0cm]{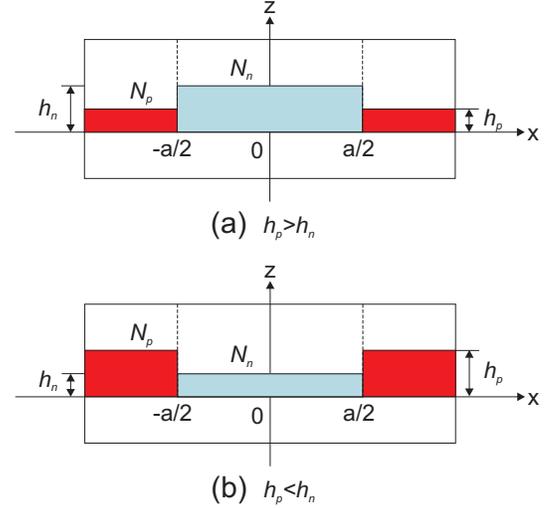}
    \caption{(Color online) Exemplary charge distributions according to Eq.~(\ref{model-rho}) (not to scale). 
             The dark-shaded rectangles are positive space charges and the light-shaded rectangle is filled
             with negative charges.}
    \label{mod-charge}
\end{center}
\end{figure}
The differences in the extensions of the positive and negative 
space charge regions originate from the different donor and acceptor concentrations and different positions of 
their energy levels in the band gap. Since the extensions  of space charge regions along the polarization direction
are by two orders of the magnitude smaller than the domain width the latter length is not expected to affect the 
mean value of the dipole density and the resulting surface potential.

To verify our understanding of the effective dipole formation we perform now an exemplary calculation of the 
potential profiles in a similar model system with asymmetric space charge zones. Two space charge distributions 
displayed in Fig.~\ref{mod-charge} are described by the charge density 
\begin{equation}
\begin{split}
\label{model-rho}
\rho_m (x_0,z_0) &= q N_p \theta\left( |x_0| -a/2 \right)\theta\left( z_0 \right)\theta\left(h_p-z_0 \right)\\
                 &- q N_n \theta\left( a/2-|x_0| \right)\theta\left( z_0 \right)\theta\left(h_n-z_0 \right)
\end{split}
\end{equation}
\noindent 
adjusted to the framework of Fig.~\ref{comp-area},where $N_n=\sigma/qh_n$ and $N_p=\sigma/qh_p$ are different 
but the total charge in the positive and negative charged areas is the same. The depths of the positive and 
negative regions are chosen, respectively, as $h_p=a/2$ and $h_n=a$ in the scheme \ref{mod-charge}(a) and as 
$h_p=a$ and $h_n=a/2$ in the scheme \ref{mod-charge}(b). 
Thanks to different extensions of the space charge regions opposite mean dipole densities $p_s$ are 
expected in the cases (a) and (b).

\begin{figure}[t]
\includegraphics[width=8.7cm]{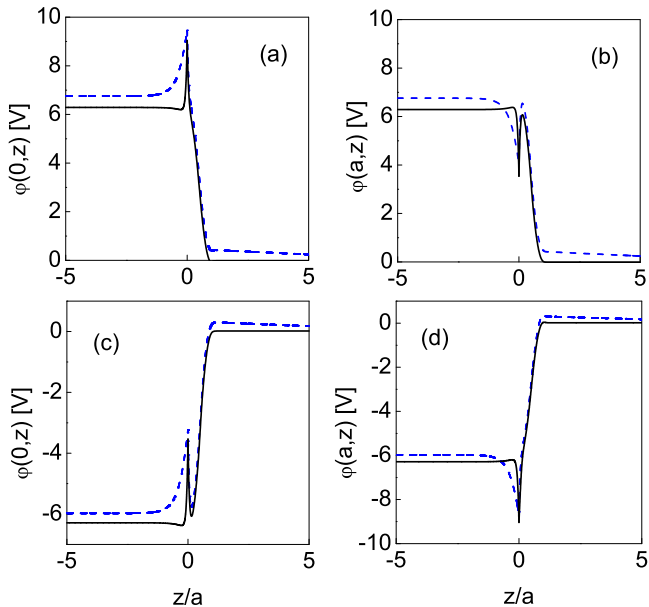}
\caption{\label{profiles-comparison}(Color online)
Electrostatic potential profiles in $z$ direction along the domain symmetry axes at $x = 0$ (a) and 
$x = a$ (b) which are produced by the model space charge distribution shown in Fig.~\ref{mod-charge}(a).
Similar potential profiles along the domain symmetry axes at $x = 0$ (c) and $x = a$ (d) produced by 
the space charge distribution in Fig.~\ref{mod-charge}(b). Solid lines represent analytical calculations
using results of Appendix B while dashed lines display FE calculations.}
\end{figure}
Potential distributions displayed in Fig.~\ref{profiles-comparison} and corresponding to the space charges 
shown in Fig.~\ref{mod-charge} were calculated, on the one hand, by using the exact analytic expressions 
(\ref{Pot-Green}), (\ref{Green-Series+}), (\ref{SymGreen+}) from Appendix B and, on the other hand, 
by means of the FE software FlexPDE. Potential profiles along the 
symmetry axes of the positive ($x=\pm a$) and negative ($x=0$) polarization domains are presented in     
Figs.~\ref{profiles-comparison}(a) and (b), respectively, for the space charge distribution in 
Fig.~\ref{mod-charge}(a) and in Figs.~\ref{profiles-comparison}(c) and (d), respectively, for the space charge 
distribution in Fig.~\ref{mod-charge}(b). 
Analytical and numerical results exhibit fair agreement revealing, however, some problems related probably 
to sharp gradients of the model space charge distributions. The results regarding the negative mean dipole 
density $p_s$, which corresponds to the positive surface potential $\varphi_s$, displayed in 
Figs.~\ref{profiles-comparison}(a,b) 
are qualitatively similar to those obtained by FE calculations in
Fig.~\ref{phiz-profile} supporting our understanding of the phenomenon of the surface potential at the
ferroelectric/dielectric interface. 
The dependence of the latter potential on the concentration and energy 
of the involved defects suggests investigation of the doping effect on this phenomenon which follows below 
in Section \ref{sec:doping}.

Though the interface at $z=0$ remains electrically neutral as a whole the effective surface charge density 
involved in the formation of the surface dipole layer can be estimated as $\sigma_d\simeq p_s/l$
where $l$ is the difference in spatial extensions between the positive and the negative space charge regions.
This length is not easy to evaluate from Figs.~\ref{charge-profiles} and \ref{Flex} where charge densities
are displayed on the logarithmic scale. From the potential profiles in Fig.~\ref{phiz-profile} it can be
estimated as $l\simeq 0.3a=3\times 10^{-8}\rm\: m$. Together with $p_s\simeq 6.6\times 10^{-10} \rm\: C/m$
from Eq.~(\ref{p_s-phi_s}) $\sigma_d\simeq 0.02 \rm\: C/m^{2}$ can be estimated which is by one order of the
magnitude smaller then the surface bound charge equal to $P_s=0.25 \rm\: C/m^{2}$.
Note that the values of the positive and negative surface charge densities evaluated separately from Fig.~\ref{Flex}
can be by one order of the magnitude larger than $\sigma_d$, namely, 
$10^9 \rm\: C/m^3$ $\times 10^{-10} \rm\: m$ $\simeq 0.1\rm\: C/m$ and thus of the order of $P_s$.

\subsection{\label{sec:Energy} Energy of a domain structure in a semiconducting ferroelectric}

Formation of the effective dipole layer and the surface potential results from a complicated balance between 
energies of the electric field, the charged defect states and charge carriers. It makes sense to evaluate the
contribution of the surface potential in this balance. To this end we use a general expression for energy density
derived in Ref. \cite{Gureev2011} for a one-dimensional domain structure in an isotropic ferrolectric with variable 
polarization which can be straightforwardly generalized to our case of a hard anisotropic ferroelectric. The 
energy density with account of screening charges of semiconductor nature reads
\begin{equation}
\label{en-density}
W=W_{field}+W_{kin}+W_{def}
\end{equation}
\noindent where the energy density of the electric field ${\bf E}$ is
\begin{equation}
\label{e-dens}
W_{field}=\frac{1}{2}\sum_{ik} \varepsilon_{ik} E_i E_k, 
\end{equation}
\noindent the density of the kinetic energy of electrons is given by     
\begin{equation}
\label{e-kin}
W_{kin}=\int_{-\infty}^{E_V} d{\cal E} Z_v({\cal E}) f({\cal E}) +
        \int_{E_C}^{\infty} d{\cal E} Z_c({\cal E}) f({\cal E}),
\end{equation}
and the energy density of charged defect states is
\begin{equation}
\label{e-def}
W_{def} = z_a N_a t_a(\varphi) E_a + z_d N_d \left( 1-t_d(\varphi) \right) E_d.
\end{equation}
\noindent Here $Z_c({\cal E})$ and $Z_v({\cal E})$ are densities of states in the conduction and valence 
bands, respectively, $f({\cal E})$ is the Fermi function, $z_a$ and $z_d$ are the acceptor and donor valences, 
respectively, $t_a(\varphi)$ and $t_d(\varphi)$ are the fractions of ionized donors and acceptors, respectively, 
and $E_a$ and $E_d$ are the donor and acceptor levels, respectively \cite{Gureev2011}.

Using Gauss' law (\ref{Gauss}) and boundary conditions the energy of the electric field (\ref{e-dens})
per one periodic unit of the stripe domain structure in Fig.~(\ref{domarray})
can be transformed to
\begin{eqnarray}
\label{e-unit}
W_{f}&=&\frac{1}{2}\int_{-a}^a dx\, \sigma _p(x,0)\,\varphi (x,0)\nonumber\\
     &+&\frac{1}{2}\int_{-a}^a dx \int_0^{\infty} dz\, \rho(x,z)\, \varphi (x,z).
\end{eqnarray}
In absence of the space charge due to charge carriers and charged defects the second term in Eq.~(\ref{e-unit}) 
disappears and this equation results in the well known expression 
\cite{LandauElectrodynamicsContinuum,Mitsui1953}
\begin{equation}
\label{e-class}
W_{f} = 0.85 \frac{P_s^2 a^2}{4\pi\varepsilon_0\sqrt{\varepsilon_a \varepsilon_c } }. 
\end{equation}

In presence of electronic charge carriers and charged defects the variation of the surface potential is reduced
by half as is seen in Fig. \ref{phix-profile}. Accordingly, the first term in Eq.~(\ref{e-unit}) is also reduced
by half with respect to the space charge-free value (\ref{e-class}). Note that a constant surface potential 
$\varphi_s$ does not contribute to this term because of the alternating surface bound charge $\sigma _p(x,0)$.
It can however contribute to the second term in Eq.~(\ref{e-unit}). In the case of intrinsic screening due to
electronic carriers only, the electrostatic potential penetrates the ferroelectric bulk to the depth of $a_s$ 
(see Fig. \ref{phiz-profile}). The corresponding contribution of the space charge in the energy (\ref{e-dens})  
is about $q\varphi_{max} n_{max} aa_s$. Since $\varphi$ is in the range of few Volts and 
$n\sim 10^{22} \rm\: m^{-3}$ at maximum (see Fig. \ref{charge-profiles}) this contribution is three orders of the 
magnitude smaller than the value (\ref{e-class}). In the presence of defects the surface potential step is formed 
so that the electrostatic potential penetrates to the depth of about $5a$ (see Fig. \ref{phiz-profile}). The
corresponding contribution to the energy still remains two orders of the magnitude smaller than (\ref{e-class})
and thus negligible.

Consider now the kinetic energy of charge carriers, Eq. (\ref{e-kin}). Since even for the peak values of the
electron and hole densities $\hbar^2 n^{2/3}/m \ll k_B T$, with $m$ the electron mass and $T$ room temperature,
the classical Boltzmann statistics applies for charge carriers. In this case the energy density (\ref{e-kin})
reduces to \cite{Gureev2011}   
\begin{equation}
\label{ekin}
W_{kin}=\left( E_F +q\varphi \right) (n-p).
\end{equation} 
\noindent Thanks to the alternating potential and carrier densities the corresponding contribution to the energy 
is positive and as small as the second term in Eq.~(\ref{e-unit}) in comparison with the value (\ref{e-class}).

The energy density of charged defect states (\ref{e-def}) does not disappear deep in the bulk of the ferroelectric
grain but saturates to the value 
\begin{equation}
\label{def-lim}
W_{def} \simeq 2N_{\left[V_{Ti}-V_O\right]^{2-}}  E_{\left[ V_{Ti}-V_O \right]^{2-}}
\end{equation}
\noindent defined by the dominating acceptor defect, the doubly ionized di-vacancies $\left[V_{Ti}-V_O\right]^{2-}$.
Since the bulk value of this defect density is about $10^{22} \rm\: m^{-3}$ (see Fig. \ref{charge-profiles})
this contribution integrated over the one unit area $a\times h$ is one order of the magnitude smaller 
than the value (\ref{e-class}).

Concluding this analysis, the energy gain due to the field screening of the semiconductor nature appears to be 
much larger than the other contributions to the energy (\ref{en-density}) including the effect of the nonzero 
surface potential. This does not mean, however, that the space charge would not have an effect on the domain 
structure if the variation of the latter were allowed. Generally, the space charge influence on domain
configurations is known to be strong \cite{Wang2008}. The results of phase-field modeling show 
that the variable periodic domain structure is remarkably modified in the presence of the semiconductor space 
charge while the surface potential at the grain remains comparable to that of the hard domain structure
considered here \cite{Zuo2013}.

\section{\label{sec:doping} Potential distributions in the extrinsically doped $\rm BaTiO_3$}

Ferroelectric perovskites are, in fact, always intentionally or unintentionally doped with various metallic 
ions widely present in the earth crust or involved in the production process \cite{Smyth_book}. Even small amounts
of them may substantially change equilibrium concentrations of the intrinsic defects emerging at sintering 
temperatures, particularly, of the oxygen vacancies. That is why the values of concentrations evaluated in 
Section \ref{subsec:defects} will change  in the presence of dopants and should be recalculated for each dopant 
type and concentration.
Likewise the value of the Fermi energy should be evaluated in each particular case. 
Nominally pure materials typically contain about $100 \rm\: ppm$, or $0.01 \rm\: mol\%$, of residual metallic 
ions \cite{WaserAPL2001}, the minimum doping value considered here. Higher intentional doping used, for example, 
for tuning of soft-hard properties of ferroelectrics \cite{MorozovJAP2010} may amount to a few per cent.

First we consider the typical case of $\rm BaTiO_3$ doped with manganese \cite{WaserAPL2001,ZhangAPL2008}, 
which may occupy the titanium site of the crystal cell, resulting in defects $\rm Mn_{Ti}^{2-}$ for the 
$\rm Mn^{2+}$ state and $\rm Mn_{Ti}^{-}$ for the $\rm Mn^{3+}$. The defect concentration calculations as 
described in Section \ref{subsec:defects} using the energy levels of different ionization states of $\rm Mn$ 
established in \cite{HagemannJACS1981,WechslerJOSA1988} show, that the major defect is singly ionized 
$\rm Mn_{Ti}^{-}$ with a transition energy of $E_{Mn_{Ti}^{-}}=1.3 \rm\: eV$. The concentration of charged $\rm Mn$ ions
\begin{equation} 
\label{Mn}
N_{Mn_{Ti}^{-}} =  \frac{N_{Mn_{Ti}} } {1 + g\exp{\left( \frac{\displaystyle E_{Mn_{Ti}}-E_F-q\varphi }
{\displaystyle k_B T} \right)}}
\end{equation} 
\noindent
with $g=4$ should be added to the right hand side of Eq.~(\ref{rho}). The values of the intrinsic defect densities
equilibrated at $1000 \rm\: K$ and corresponding room temperature Fermi energies are self-consistently calculated 
for different doping concentrations $\rm N_{Mn_{Ti}}$ by the procedure developed in Ref. \cite{ErhartJAP2008} and 
shown in Table II.

\begin{figure}[t]
\includegraphics[width=7.4cm]{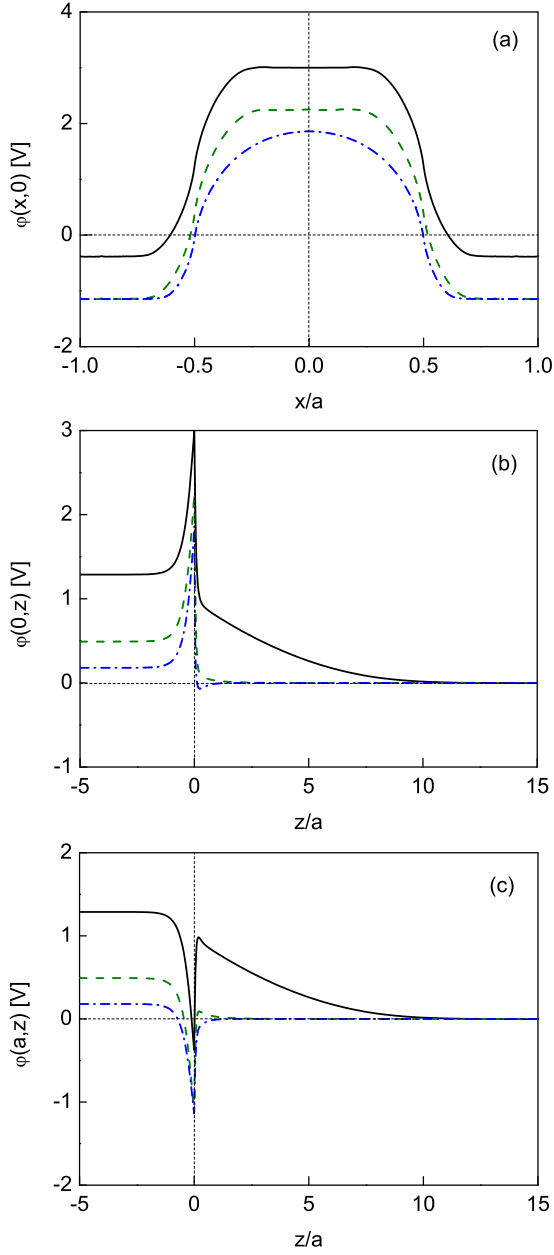}
\caption{\label{profiles-doped} (Color online)
Electrostatic potential profile in $x$ direction along the interface at $z = 0$ (a) and in $z$ direction along the 
domain symmetry axes at $x = 0$ (b) and $x = a$ (c). Solid, dashed and dash-dotted lines show the potential
distributions for doping of $0.01$, $0.1$ and $1 \rm\: mol\%$, respectively.}
\end{figure}

Solving
Eqs.~(\ref{Gauss},\ref{rho},\ref{n-density},\ref{p-density},\ref{VO-density},\ref{Ti-density},\ref{TiO-density},\ref{Mn}) 
with input parameters from Table~\ref{Materialparameters2} by means of FlexPDE results in the electrostatic potential profiles 
displayed in Fig.~\ref{profiles-doped}. Some features distinguish these profiles from those of the intrinsically doped
material in Figs.~\ref{phix-profile} and \ref{phiz-profile}. The difference between the minima of the $0.01\rm\: mol\%$-line 
and the $0.1\rm\: mol\%$-line in Fig.~\ref{profiles-doped}(a) is about $0.75\rm\: V$. Raising the manganese level by
one order of the magnitude to one mole percent does not change the value of the potential in the middle of the negative
domain at $x=a$. The potential in the middle of the positive domain at $x=0$, on the other hand, decreases further 
remarkably with rising doping level.

\begin{table}[t]
\caption{\label{Materialparameters2}Defect densities and Fermi energy of $\rm Mn$-doped $\rm BaTiO_3$. }
\begin{center}
\begin{ruledtabular}
\renewcommand{\arraystretch}{1.5}
\begin{tabular}{cccc}
& $0.01 \rm\: mol\%$ & $0.1 \rm\: mol\%$ & $1 \rm\: mol\%$ \vspace{2pt}\\
\hline
$E_F\,[\rm\:eV]$ & 0.390 & 1.148  & 1.148 \\
$N_{V_{O}}\,[\rm\:m^{-3}]$ & $5.887 \cdot 10^{20} $ & $5.811 \cdot 10^{21} $ &  $5.600 \cdot 10^{22} $\\
$N_{V_{Ti}}\,[\rm\:m^{-3}]$ & $3.612 \cdot 10^{20} $ & $3.713 \cdot 10^{18} $ &  $4.017 \cdot 10^{16} $\\
$N_{[V_{Ti}-V_{O}]}\,[\rm\:m^{-3}]$ & $2.825 \cdot 10^{21} $ & $2.863 \cdot 10^{20} $ &  $2.972 \cdot 10^{19} $\\
$N_{Mn_{Ti}}\,[\rm\:m^{-3}]$ & $1.563 \cdot 10^{24} $ & $1.563 \cdot 10^{25} $ &  $1.563 \cdot 10^{26} $ 
\end{tabular}
\end{ruledtabular}
\vspace{0.5cm}
\end{center}
\end{table}

Potential profiles along the $z$-direction for the lowest doping of $0.01\rm\: mol\%$ remind of the case of 
intrinsic defects (dash-dotted lines in Fig.~\ref{phiz-profile}), though with substantially enhanced penetration 
depth of the electric field. The profiles corresponding to higher doping become, in contrast, substantially different. 
Similarly, the mean value of the potential, $\varphi_s=1.287\rm\: V$, at the interface $z = 0$ for  
$0.01 \rm\: mol\%$ doping is not very different from the value of $1.345 \rm\: V$ in the case of intrinsic doping. 
The values of $\varphi_s=0.492\rm\: V$ and  $0.180\rm\: V$ for $0.1$ and $1 \rm\: mol\%$ doping, respectively,
are, however, notably different. 

\begin{figure}[t]
\includegraphics[width=7.4cm]{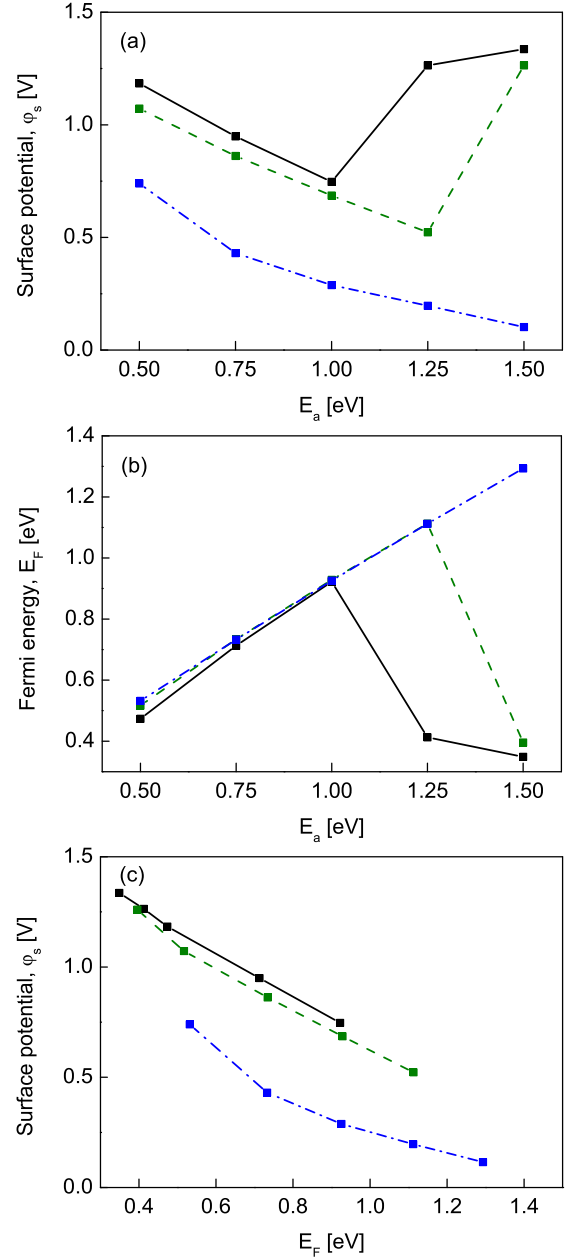}
\caption{\label{doping-deps} (Color online)
Surface potential at the grain boundary (a) and the Fermi energy (b) as functions of the acceptor defect energy 
with respect to the top of the valence band for $0.01$ (solid line), $0.1$ (dashed line) and $1 \rm\: mol\%$ 
(dash-dotted line) doping, respectively. Similarly indicated dependences of the surface potential on the 
Fermi energy (c) for different doping.}
\end{figure}

The electric potential profile in the $z$-direction of the intrinsically doped material exhibits a maximum at about 
$z=0.2a$ in front of the negatively charged interface ($x = a$). A similar maximum can be seen at $z=0.2a$ for the 
$0.01$ and $0.1 \rm\: mol\%$ doped sample but vanishes for higher doping. The penetration depth of the electric
potential is rather large for the least doped sample. It takes about ten times the domain width 
$a = 100\rm\:nm$ to reach zero. In the higher doped samples it occurs at a much shorter distance. 

Considering the strong effect of even medium doping on the surface potential at the ferroelectric grain it is 
interesting to investigate the influence of different possible dopants on this potential. To this end all the above 
calculations of energy, charge and potential distributions including the equilibration with intrinsic defects at 
sintering temperature were repeated adopting different doping concentrations of hypothetical, simply ionized 
acceptor defects with different energy level positions in the band gap. In Fig.~\ref{doping-deps} the dependencies 
of the surface potential and the Fermi energy on the defect energy $E_a$ for different doping levels are presented. 
The surface potential is found to be tunable by acceptor doping in a wide range from $0.1$ to $1.3 \rm\: V$ and is 
apparently correlated with the Fermi level position exhibiting virtually linear descending dependence on the latter
(Fig.~\ref{doping-deps}(c)). Interestingly, higher extrinsic doping concentrations depress the surface potential by 
reducing the effect of the intrinsic defects $V_{Ti}$ and $\left[V_{Ti}-V_O\right]$ as is clearly seen from 
Table~\ref{Materialparameters2}.

\section{\label{sec:conclusions}Conclusions}

Very high variation of the electrostatic potential between alternatively charged polarization domain boundaries in a 
ferroelectric domain array makes it necessary to account for the local electronic band bending at the typical scale of 
the domain width. In the current work this problem has been numerically treated within a two-dimensional semiconductor 
model of a ferroelectric grain supported by the analytic treatment of the linear dielectric model. In contrast to 
expectations \cite{WatanabeFerro2008,Kaku2009} the nonlinear screening of the depolarization fields by formation of the
electronic and the defect space charges due to the band bending cannot explain the reduction of the potential variations 
by orders of magnitude. Fig.~\ref{phix-profile} shows the decrease of the potential magnitude by approximately one half 
with respect to the ideal dielectric value \cite{LandauElectrodynamicsContinuum}. This means that the surface potential
variations observed in \cite{Kaku2009} are most probably of different nature, or that the potential variations due to
polarization are strongly compensated by other physical mechanisms mentioned in \cite{Kaku2009}, for example, by 
field-driven oxygen vacancy migration \cite{Genenko-agingPRB2007,GenenkoPRB2008}.

Another conclusion following from the analysis of nonlinear field screening within the semiconductor model is that the 
maximum amplitude of local electric depolarization fields in ferroelectrics appears to be determined not by the polarization 
$P_s$ and the permittivity but rather by the electronic band structure because the typically very large depolarization fields 
of the magnitude $\sim P_s/\varepsilon_0 \varepsilon_c$ are limited by the value about $E_g/qa$ due to screening of 
semiconducting nature. In an unpoled ferroelectric the characteristic length $a$ is given by the typical domain width, 
in the highly poled ferroelectric ceramic by the typical size of the poled region, say, the grain size. The latter
limitation entails a reduction of the remanent depolarization fields in polycrystalline material with larger grain 
size and, therefore, a decrease of the internal bias field characterizing aging in the poled state by charge 
migration; a phenomenon observed in experiments \cite{GenenkoPRB2009}.

Taking into account typical intrinsic defects which develop during the high temperature processing of ferroelectric 
ceramics reveals unexpected features of nonlinear field screening, namely, the formation of an effective dipole layer 
at the ferroelectric grain boundary due to unbalanced space charge regions in front of differently charged domain 
boundaries. This dipole layer results in a surface electrostatic potential at the grain boundary which can be of either 
sign and on the order of $1\rm\: V$. Such a potential may have a dramatic impact on both ionic and electronic transport 
in ferroelectric ceramics by modifying the potential barriers for charge carriers at the grain boundaries. The
magnitude of the obtained surface potential appears to be very sensitive to low doping levels of about 0.01\% and is 
generally reduced at higher doping levels remaining nevertheless remarkably large. Particularly, acceptor or donor 
doping allows fine tuning of this surface potential between roughly $-1$ and $1\rm\: V$.

\begin{acknowledgments} 
This work was supported by the Deutsche Forschungsgemeinschaft through the Sonderforschungsbereich 595 "Electrical 
Fatigue in Functional Materials".
\end{acknowledgments}

\appendix
\section*{Appendix A. Potential of a straight charged line parallel 
to a boundary between anisotropic and isotropic semi-spaces}

Consider a semi-space $z>0$ occupied  by an anisotropic dielectric medium characterized by the tensor of dielectric
permittivity $\hat\varepsilon=\varepsilon_0 \hat\varepsilon_f$ with the relative permittivity given by 
Eq.~(\ref{etensor}). The lower semi-space $z<0$ is occupied by an isotropic dielectric medium with $\hat\varepsilon=\varepsilon_0 \varepsilon_d \hat 1$, with $\hat 1$ the unit tensor.

A straight charged line  with a charge density $\tau $ per unit length oriented parallel to the 
 $y-$axis and, thus, to the boundary between the two media, $z=0$, crosses the $(x,z)$ plane at the point
$(x_0,z_0)$ with $z_0>0$. Thanks to the translational symmetry along the $y-$axis all potentials and 
fields depend only on $x$ and $z$.

For the charge-free area $z<0$ the Laplace equation for the electrostatic potential $\varphi$ applies:
\begin{equation}
\label{Laplace}
\partial_x^2\varphi + \partial_z^2\varphi  =0 
\end{equation}
For the area $z>0$ the Poisson equation     
\begin{equation}
\label{Poisson}
\varepsilon_a \partial_x^2\varphi + \varepsilon_c \partial_z^2\varphi  = -(\tau/\varepsilon_0) 
\delta({\bf r}-{\bf r}_0) 
\end{equation}
is valid with the two-dimensional Dirac $\delta-$function and radius-vectors 
${\bf r}=(x,z)$ and ${\bf r}_0=(x_0,z_0)$.
Boundary conditions at the interface $z=0$ are 
\begin{eqnarray}
\label{BCPotential}
\varphi|_{z=-0} &  = & \varphi|_{z=+0}\\
\label{BCDisplacement}
\varepsilon_d \partial_z \varphi|_{z=-0} & = & \varepsilon_c \partial_z \varphi|_{z=+0}. 
\end{eqnarray}  


The ansatz for the potential in the area $z<0$ which satisfies equations (\ref{Laplace},\ref{Poisson}) 
as well as boundary conditions (\ref{BCPotential},\ref{BCDisplacement}) reads 
\cite{LandauElectrodynamicsContinuum,jackson75classical} 
\begin{equation}
\label{Ansatz-}
\varphi = -\frac{\tau'}{4\pi\varepsilon_0} \ln{\left(\left|\frac{{\bf r}-{\bf r}_2 }{a}\right|^2 \right)} + A
\end{equation}
where  ${\bf r}_2=(x_0,z_2)$ with $z_2=z_0\sqrt{\varepsilon_a/\varepsilon_c}$, $a$ is some characteristic length 
and A is a constant. For the area $z\geq 0$ the appropriate potential reads
\begin{equation}
\label{Ansatz+}
\varphi = -\frac{\tilde\tau}{4\pi\varepsilon_0} 
\ln{ \left( \left| \frac{\tilde{\bf r}-\tilde{\bf r}_0}{a} \right|^2 \right) }
-\frac{\tau''}{4\pi\varepsilon_0} 
\ln{ \left( \left| \frac{\tilde{\bf r}-\tilde{\bf r}_1}{a} \right|^2 \right) }
\end{equation}
where $\tilde \tau =\tau/\sqrt{\varepsilon_a \varepsilon_c}\,$, $\tilde{\bf r}=(\tilde x,\tilde z)$ 
with $\tilde x=x/\sqrt{\varepsilon_a },\, \tilde z = z/\sqrt{\varepsilon_c}\,$, 
$\tilde {\bf r}_0=(\tilde x_0,\tilde z_0)$ 
with $\tilde x_0=x_0/\sqrt{\varepsilon_a },\, \tilde z_0 = z_0/\sqrt{\varepsilon_c}$ and 
$\tilde {\bf r}_1=(\tilde x_0,-\tilde z_0)$. The constants $A,\,\tau'$ and $\tau''$ can be determined
from the boundary conditions. By substituting the ansatz forms (\ref{Ansatz-},\ref{Ansatz+}) into Eqs.
(\ref{BCPotential},\ref{BCDisplacement}) one finds
\begin{eqnarray}
A & = & \frac{\tau }{4\pi\varepsilon_0} \frac{2 \ln{\varepsilon_a}}{\sqrt{\varepsilon_a \varepsilon_c}+
\varepsilon_d} \nonumber\\
\tau' & = & \frac{2 \tau }{\sqrt{\varepsilon_a \varepsilon_c}+\varepsilon_d} \nonumber\\
\label{constants}
\tau'' & = & \frac{\tau}{\sqrt{\varepsilon_a \varepsilon_c}} 
\frac{\sqrt{\varepsilon_a \varepsilon_c}-\varepsilon_d}{\sqrt{\varepsilon_a \varepsilon_c}+\varepsilon_d}. 
\end{eqnarray}  

For the special case of a charged line located right at the interface, $z_0=0$, the potential acquires 
the form 
\begin{equation}
\label{linepot-}
\varphi = -\frac{\tau}{4\pi\varepsilon_0} \frac{2}{\sqrt{\varepsilon_a \varepsilon_c}+\varepsilon_d}
 \ln{\left[ \frac{(x-x_0)^2+z^2}{a^2\varepsilon_a} \right]}
\end{equation}
for $z<0$, and 
\begin{equation}
\label{linepot+}
\varphi = -\frac{\tau}{4\pi\varepsilon_0} \frac{2}{\sqrt{\varepsilon_a \varepsilon_c}+\varepsilon_d}
 \ln{\left[ \frac{(x-x_0)^2}{a^2\varepsilon_a}+\frac{z^2}{a^2\varepsilon_c} \right]}
\end{equation}
for $z\geq 0$.

\section*{Appendix B. Electric field produced by an arbitrary space charge within a stripe domain array}

Here we study analytically, for a system introduced in Appendix A, a linear problem of two-dimensional array 
of domains infinite in the positive $z-$direction, periodic in the $x-$direction and cut by the surface, 
$z=0$, perpendicular to the direction of spontaneous polarization in domains. Boundary conditions 
(\ref{BCPotential},\ref{BCDisplacement}) are used.
First we calculate the field ${\bf E}^0(x,z)$ of the domain array alone without any free charges in the system. 
Then we formally solve equation (\ref{Gauss}) and find the total electric field ${\bf E}(x,z)$ for an arbitrary 
right-hand side.

The bound charge density of the domain faces with a period $a$ along the $x$-axis is represented by an alternating
function \cite{LandauElectrodynamicsContinuum}
\begin{align}
\label{face-charge}
\rho_b(x,z)=\sigma\delta(z)\sum_{n}   
(-1)^n \theta\left(\frac{a}{2}-an+x\right)
\theta\left(\frac{a}{2}+an-x\right)
\end{align}
\noindent where $\delta (z)$ and $\theta  (x)$ are the Dirac $\delta $-function and the Heaviside unit
step function, respectively. The electrostatic potential induced by this bound charge is given by
the expression
\begin{equation}
\begin{split}
\label{face-potential-} \varphi_b(x,z) & =
\frac{-1}{2\pi \varepsilon_0(\sqrt{\varepsilon_a\varepsilon_c}+\varepsilon_d)}
\int_{-\infty}^{\infty}dx_0 \int_{-\infty}^{\infty}dz_0\, \rho_b(x_0,z_0)\\
&\times \ln\left[\frac{\left(x-x_0\right)^2 + \left(z-z_0\right)^2}{a^2\varepsilon_a} \right]
\end{split}
\end{equation}
\noindent in the area $z<0$ and by the expression 
\begin{equation}
\begin{split}
\label{face-potential+} \varphi_b(x,z) &= \frac{-1}{2\pi \varepsilon_0
(\sqrt{\varepsilon_a\varepsilon_c}+\varepsilon_d)}
\int_{-\infty}^{\infty}dx_0 \int_{-\infty}^{\infty}dz_0\, \rho_b(x_0,z_0)\\
& \times\ln\left[\frac{\left(x-x_0\right)^2}{a^2\varepsilon_a} +
\frac{\left(z-z_0\right)^2}{a^2\varepsilon_c} \right]
\end{split}
\end{equation}
\noindent in the area $z\geq 0$. The formulas (\ref{face-potential-},\ref{face-potential+}) are obtained by 
a simple superposition of the potentials generated by straight parallel charged lines located at the grain 
boundary $z=0$ between the isotropic and the anisotropic media  given by Eqs.~(\ref{linepot-},\ref{linepot+})
in Appendix A.

The $z$-component of the electric field created by the bound charge, ${\bf E}^0=-\nabla \varphi_b$,  may be directly
calculated by substitution of Eq.~(\ref{face-charge}) into Eqs.~(\ref{face-potential-},\ref{face-potential+}),
differentiation and subsequent summation \cite{prudnikov86integrals} which results in the form 
\begin{equation}
\label{E0z-}
E^0_z(x,z)=\frac{2\sigma}{\pi\varepsilon_0}
\frac{1}{(\sqrt{\varepsilon_a\varepsilon_c} + \varepsilon_d )}
\arctan{\left[\frac{\cos(\pi x/a)}{\sinh{(\pi z/a)}}\right]}
\end{equation}
\noindent valid inside the dielectric medium ($z< 0$), and in the form
\begin{equation}
\begin{split}
\label{E0z+}
E^0_z(x,z) &= \frac{2\sigma}{\pi\varepsilon_0}\sqrt{\frac{\varepsilon_a}{\varepsilon_c}}
\frac{1}{(\sqrt{\varepsilon_a\varepsilon_c} + \varepsilon_d )}\\
& \times \arctan{\left[\frac{\cos(\pi x/a)}{\sinh(\sqrt{\varepsilon_a/\varepsilon_c}\,\pi z/a)}\right]}
\end{split}
\end{equation}
\noindent valid inside the ferroelectric medium ($z\geq 0$) \cite{GenenkoFerro2008}. 

Direct calculation of the other field component, $E^0_x=-\partial_x\varphi_b$, is more complicated because of slow
convergence of the respective series. Instead, $E^0_x$ may be calculated for $z\neq 0$ from Gauss' law 
$\nabla{\bf E}^0=0$, taking into account that, from the bilateral symmetry of the problem (see Fig.~\ref{domarray}),
$E^0_x(0,z)=E^0_x(\pm a,z)=0$. Proceeding with integration of the latter Gauss' equation over distance along the 
$x$-axis and using the aforementioned boundary conditions one finds the form
\begin{equation}
\label{E0x-}
E^0_x(x,z)=\frac{\sigma}{\pi\varepsilon_0 (\sqrt{\varepsilon_a\varepsilon_c} + \varepsilon_d )}
\ln{\left[\frac{\cosh(\pi z/a)+\sin(\pi x/a)}
{\cosh(\pi z/a)-\sin(\pi x/a)}\right]}
\end{equation}

\noindent valid for $z<0$ and
\begin{equation}
\begin{split}
\label{E0x+}
E^0_x(x,z) &= \frac{\sigma}{\pi\varepsilon_0 (\sqrt{\varepsilon_a\varepsilon_c} + \varepsilon_d )}\\
& \times \ln{\left[\frac{\cosh(\sqrt{\varepsilon_a/\varepsilon_c}\,\pi z/a)+\sin(\pi x/a)}
{\cosh(\sqrt{\varepsilon_a/\varepsilon_c}\,\pi z/a)-\sin(\pi x/a)}\right]}
\end{split}
\end{equation}
\noindent valid for $z\geq 0$.
Both field components exhibit periodic dependence along the $x$-axis, as expected from the periodic domain arrangement, 
and exponential decay at large distance from the charged surface $|z|\gg a$, as expected from the previous finite
element simulations \cite{GenenkoFerro2008}. The closed forms Eqs.~(\ref{E0z-}-\ref{E0x+}) are numerically
identical to the solutions in terms of Fourier series given in \cite{LandauElectrodynamicsContinuum,Fedosov1976}
and reduce to the previously derived expressions for the isotropic case \cite{Genenko-agingPRB2007}.

In the presence of a space charge density $\rho_i(x,z)$  in the area $z>0$, the total electric field in the considered 
linear problem may be conveniently decomposed as ${\bf E}={\bf E}^0+{\bf E}^i$, where the field ${\bf E}^0$ is determined 
by the bound charge of the domains, $\rho_b(x,z)$, and the field ${\bf E}^i$ is generated by the free charge distribution
$\rho_i(x,z)$. Thanks to the periodicity and the bilateral symmetry of the boundary conditions, the region $-a<x<a$
can be used as a repetitive basic unit of the system. To get a full description of the electric field under these
circumstances, it is sufficient to construct Green's function of the symmetrical Neumann problem in the  
region, $G_s(x,z|x_0,z_0)$, so that the electrostatic potential induced by the charge density $\rho_i(x,z)$
can be presented in a form \cite{jackson75classical}
\begin{equation}
\label{Pot-Green}
\varphi_i(x,z)=\int_{0}^{a}dx_0 \int_0^{\infty}dz_0\,
\rho(x_0,z_0) G_s(x,z|x_0,z_0),
\end{equation}
followed by the field expression ${\bf E}^i=-\nabla \varphi_i$.

Green's function satisfies the Laplace equation in the area $z<0$ and the equation
\begin{multline}
\label{Eq-Green}
\varepsilon_0\left(\varepsilon_a \partial_x^2 +\varepsilon_c \partial_z^2\right) G_s(x,z|x_0,z_0) =\\
 -\delta (z-z_0) \left[\delta(x-x_0) + \delta (x+x_0)\right]
\end{multline}
\noindent in the area $z\geq 0$ with boundary conditions $\partial_x G_s(x=\pm a ,z|x_0,z_0)=0$. The latter requirement 
is a consequence of the constraint $E_{x}(\pm a,z)=0$ inherent to the chosen domain arrangement. Boundary conditions for 
the electrostatic potential on the interface between the two media at $z=0$, 
Eqs.~(\ref{BCPotential},\ref{BCDisplacement}), impose two additional boundary conditions on Green's 
function
\begin{align}
\label{Green-Boundaries}
G_s(x,-0|x_0,z_0) &= G_s(x,+0|x_0,z_0)\nonumber\\
\varepsilon_d \partial_z G_s(x,-0|x_0,z_0) &=\varepsilon_c \partial_z G_s(x,+0|x_0,z_0).
\end{align}

By using the fundamental solution of the 2D Poisson equation \cite{jackson75classical} (see Appendix A) and taking into 
account the periodicity of the problem the solution of Eq.~(\ref{Eq-Green}) may be reduced to summation of the series
\begin{multline}
\label{Green-Series-}
G_s(x,z|x_0,z_0) =  -\frac{1}{2\pi \varepsilon_0(\sqrt{\varepsilon_a\varepsilon_c} + \varepsilon_d)}\\ 
\times\sum_{n}   
\left\{\ln\left[ \frac{(x-x_0-2an)^2 + (z-z_0\sqrt{\varepsilon_a/\varepsilon_c})^2}{a^2\varepsilon_a} 
\right]\right\} \\
+  (x_0 \rightarrow -x_0)
\end{multline}
\noindent for the area $z<0$ and
\begin{multline}
\label{Green-Series+}
G_s(x,z|x_0,z_0)=-\frac{1}{4\pi \varepsilon_0\sqrt{\varepsilon_a\varepsilon_c}}\\
\times\sum_{n} \left\{\ln\left[ \frac{(x-x_0-2an)^2}{a^2\varepsilon_a}+
\frac{(z-z_0)^2}{a^2\varepsilon_c}\right]\right.\\
\left.
+\frac{\sqrt{\varepsilon_a\varepsilon_c} - \varepsilon_d}
{\sqrt{\varepsilon_a\varepsilon_c} + \varepsilon_d}
\ln\left[ \frac{(x-x_0-2an)^2}{a^2\varepsilon_a}+
\frac{(z+z_0)^2}{a^2\varepsilon_c}\right]\right\}\\
+ (x_0 \rightarrow -x_0)
\end{multline}
\noindent for the area $z\geq 0$.

Because of slow convergence of these series it is more convenient to perform summation for the derivatives 
$\partial_x G_s$ and $\partial_z G_s$ and then to restore the function $G_s$ itself by integration using boundary 
conditions. This leads eventually to 
\begin{multline}
\label{SymGreen-}
G_s(x,z|x_0,z_0) = -\frac{1}{2\pi\varepsilon_0(\sqrt{\varepsilon_a\varepsilon_c} + \varepsilon_d)}\\
\times\ln\left[ \cosh\frac{\pi(z-z_0\sqrt{\varepsilon_a/\varepsilon_c})}{a} -\cos\frac{\pi(x-x_0)}{a} \right]\\
+(x_0 \rightarrow -x_0)
\end{multline}
\noindent for the area $z<0$ and
\begin{multline}
\label{SymGreen+}
G_s(x,z|x_0,z_0)=-\frac{1}{4\pi\varepsilon_0 \sqrt{\varepsilon_a\varepsilon_c}}\\
\times\left\{\ln\left[ \cosh{\left(\sqrt{\frac{\varepsilon_a}{\varepsilon_c}}\frac{\pi(z-z_0)}{a}\right)} -
\cos\frac{\pi(x-x_0)}{a} \right]\right.\\
\left.+\frac{\sqrt{\varepsilon_a\varepsilon_c} - \varepsilon_d}{\sqrt{\varepsilon_a\varepsilon_c} + \varepsilon_d}
\ln\left[ \cosh{\left(\sqrt{\frac{\varepsilon_a}{\varepsilon_c}}\frac{\pi(z+z_0)}{a}\right)}\right.\right.\\ 
\left.\left. -\cos\frac{\pi(x-x_0)}{a} \right]\right\} + (x_0 \rightarrow -x_0)
\end{multline}
\noindent for the area $z\geq 0$, which is periodic, bilaterally symmetric and satisfies the proper boundary conditions. 
This solution reduces also to the previously derived one in the isotropic limiting case \cite{Genenko-agingPRB2007}.

\bibliographystyle{plain}
\bibliography{apssamp}

\end{document}